\documentclass[sigconf,nonacm=true,10pt]{acmart}

\newtheorem{principle}{Rule}

\begin{document}

\title{Customizing Trusted AI Accelerators for Efficient Privacy-Preserving Machine Learning}

\author{Peichen Xie}
\affiliation{Peking University}
\author{Xuanle Ren}
\affiliation{Alibaba Group}
\author{Guangyu Sun}
\affiliation{Peking University}

\begin{abstract}
The use of trusted hardware has become a promising solution to enable privacy-preserving machine learning.
In particular, users can upload their private data and models to a hardware-enforced trusted execution environment (e.g. an \textit{enclave} in Intel SGX-enabled CPUs) and run machine learning tasks in it with confidentiality and integrity guaranteed.
To improve performance, AI accelerators have been widely employed for modern machine learning tasks.
However, how to protect privacy on an AI accelerator remains an open question.
To address this question, we propose a solution for efficient privacy-preserving machine learning based on an unmodified trusted CPU and a customized trusted AI accelerator.
We carefully leverage cryptographic primitives to establish trust and protect the channel between the CPU and the accelerator.
As a case study, we demonstrate our solution based on the open-source versatile tensor accelerator.
The result of evaluation shows that the proposed solution provides efficient privacy-preserving machine learning at a small design cost and moderate performance overhead.

\end{abstract}

\maketitle

\section{Introduction}

Privacy is a key issue in many machine learning applications such as machine learning as a service, federated learning, inference at the edge, etc. 
Generally, we formulate a machine learning task by $$\mathrm{result}=f(\mathrm{model}, \mathrm{data})$$
where the model and the data can be provided by two or more parties. 
The term \textit{privacy-preserving machine learning} is to evaluate $f(\mathrm{model}, \mathrm{data})$ without disclosing their private model/data.

To this end, numerous recent studies regard it as a secure multi-party computation (MPC) problem and propose cryptography-based solutions.
For example, homomorphic encryption, garbled circuits and secret sharing are leveraged for privacy-preserving inference \cite{gilad-bachrach16,juvekar18} and privacy-preserving training \cite{mohassel17,rouhani18}.
However, cryptography-based solutions are very inefficient because of the intensive computation and communication of homomorphic encryption and MPC protocols.
In addition, these solutions lack versatility because the operations supported by homomorphic encryption and MPC protocols are limited.

Alternatively, trusted execution environment (TEE) is a promising approach to efficient privacy-preserving machine learning.
Lately, CPU designers have integrated trusted computing components into CPUs which enables hardware-based trusted execution environments.
Based on trusted CPUs (e.g. Intel's SGX-enabled CPUs), the model/data owners can upload their private model/data to a secure \textit{enclave} via encrypted communication channels.
The enclave is secured such that only verified trusted software can access and decrypt the model and data.
In the enclave, the CPU evaluates $f(\mathrm{model}, \mathrm{data})$ over the decrypted model and data.
Therefore, compared with cryptography-based solutions, such approaches are more versatile and can achieve privacy-preserving machine learning with lower overhead \cite{ohrimenko16,hunt18}.




However, the computational power of CPUs is still insufficient for large-scale ML models such as deep neural networks.
To improve performance, both academia and industry have designed various dedicated AI accelerators such as Eyeriss\cite{chen16}, TPUs\cite{jouppi17}, VTA\cite{moreau19}, etc.
Although AI accelerators can help handle such workloads, employing AI accelerators as well as protecting privacy remains an open question (Section \ref{challenges}).
For example, Volos et al. \cite{volos18} have proposed enabling TEEs on GPUs, but this solution relies on the assumption that recent server-class GPUs have trusted device memory (e.g. on-package HBM).
This assumption, however, is not appropriate for many AI accelerators, which use off-package untrusted memory. 
Jiang et al. \cite{jang19} propose to customize the CPU hardware to provide secure I/O paths to the GPU.
However, it omits physical attacks (e.g. eavesdropping on the bus), and modifying the architecture/behavior of modern CPUs may also be difficult for most companies and institutions.

In contrast, in this paper, we propose a simple but effective solution for efficient privacy-preserving machine learning by customizing trusted AI accelerators without modifying the CPU.
Specifically, we customize the accelerator by adding a security interface and a crypto engine to the AI accelerator; the CPU, the core logic of the accelerator and the buses remain unchanged.
To ensure confidentiality and integrity, we first leverage the newly-added cryptographic primitives to establish trust and exchange a symmetric key between the CPU enclave and the AI accelerator.
Then, we carefully protect all communication channels between the CPU enclave and the accelerator using this symmetric key.
Since the key is only held by the CPU enclave and the accelerator, anyone without the key cannot obtain or tamper with the code and data.
As a result, we can offload the computation to the accelerator securely, and leverage the accelerator to evaluate $f(\mathrm{model}, \mathrm{data})$ efficiently.

As a case study, we demonstrate the implementation of our solution on an open-source AI accelerator (Section \ref{casestudy}).
The result shows that our solution can utilize the high computational power of the AI accelerator with major hardware/software designs (such as the IP core design, the instruction set architecture and the compiler) unchanged.
Therefore, this solution is suitable for the scenarios where AI accelerators are customizable but CPUs/GPUs are not. 

In summary, the contributions of our paper are as follows:
\begin{itemize}
\item We propose the solution for efficient privacy-preserving machine learning by customizing trusted AI accelerators and extending TEEs to AI accelerators.
\item We propose the methods to secure the code and data used by the customized AI accelerator. Particularly, the methods are effective even if the AI accelerator does not have trusted device memory. 
\item We evaluate the implementation of our solution on VTA (an open-source AI accelerator) as a case study, and  analyze the performance impact with cycle-accurate register-transfer-level simulation.
\item We analyze the performance overhead of our solution and present several ways to reduce the overhead.
\end{itemize}

\section{Background}
\subsection{Trusted Execution Environment}

A trusted execution environment (TEE) guarantees code and data in it are isolated and protected from the outside environment, including unauthorized users, attackers, system administrators and even the operating system and the hypervisor.
Various implementations of the TEE have been proposed by industry and academia recently, e.g. Intel SGX \cite{intel_2016}, ARM TrustZone \cite{arm_2009}, Keystone \cite{lee_2020}, based on their respective CPUs.
In this paper, we use Intel SGX as the CPU-side TEE because it provides comprehensive protection and it is industry-ready.
In addition, because SGX has a minimal trusted computing base (TCB) \cite{costan16}, our solution can be adapted to other CPUs.

SGX-enabled CPUs can protect a specific memory region (i.e. processor’s reserved memory) from all unauthorized memory accesses and DMA accesses thanks to the integration of the uncore (including the memory controller and the I/O controller).
Besides, the SGX-enabled CPU has integrated a memory encryption engine in order to protect this memory region against physical attacks.
Based on these security features, Intel provides secure containers called \textit{enclaves} that hold private data and code in the processor’s reserved memory.

For privacy-preserving machine learning, the typical solution based on SGX is to have different parties upload their confidential data/model into an enclave over an encrypted channel, attest the software running in the enclave and finally receive the encrypted result \cite{ohrimenko16,hunt18}.
However, using trusted CPUs, the performance is restricted by the CPU performance, which is not sufficient for modern machine learning tasks (e.g. training a deep neural network).

\subsection{Extending TEEs to Accelerators}
\label{challenges}
Recent studies have attempted several ways to leverage hardware accelerators to improve performance while ensuring data privacy and security.
In this part, we will give a brief introduction to them and explain why it is still challenging to reach that goal.

Trusted I/O is a generic way to establish a trusted and secure I/O path to a target I/O device.
Regretfully, SGX lacks support for generic trusted I/O; SGX does not support running privileged code in an enclave, either, which is important for managing I/O devices.
Although recent works \cite{weiser_2017,liang_2020} have proposed trusted path architectures for SGX, their methods compromise the security of SGX (e.g. they do not consider physical attacks). 

Compared with the I/O devices without device memory (e.g. a keyboard), accelerators have a larger attack surface because of their own code and data stored in memory.
Thus, it is difficult to use accelerators securely by just applying generic trusted I/O.
Instead, designing a trusted path to the accelerator specifically is a more promising way.

HIX \cite{jang19} is a hardware/software architecture to protect the I/O path between user software and an unmodified commodity GPU.
To achieve this, HIX changes the CPU's hardware architecture in order to 1) provide a specific enclave for the GPU driver and 2) protect MMIO accesses to the GPU.
As a result, it provides protection against privileged software.
However, HIX assumes the whole graphics card (including the GPU memory) and the whole hardware system (including buses) are trusted.
Therefore, the assumption suggests that HIX cannot prevent physical attacks (such as eavesdropping on the bus) yet.


Instead of customizing the CPU, Graviton \cite{volos18} modifies the GPU to support TEEs on GPUs. 
In particular, it manages to provide secure ``contexts'' on the multi-kernel GPU by customizing the GPU's command processor and addresses the problem of physical snooping attacks.
The design is based on the assumption that the GPU has on-package memory (e.g. HBM) which is within the trust boundary.
This assumption is appropriate for modern server-class GPUs.
However, many AI accelerators use off-package memory which is commonly assumed to be untrusted, so the design of Graviton cannot be directly adapted to AI accelerators.

Slalom \cite{tramer19} provides a novel framework to outsource matrix multiplication to an untrusted hardware accelerator, in order to accelerate deep neural network (DNN) inference, based on existing trusted CPUs.
To ensure privacy and integrity, it leverages two cryptographic schemes, i.e. additive stream cipher and Freivalds' algorithm.
However, Slalom has the following limitations: 
1) It omits the model privacy with respect to the server. 
2) As the authors acknowledge, applying Slalom to DNN training is hard because of the limitation of algorithm.
3) In terms of performance, Slalom requires a pre-processing phase but regretfully, the paper has reported neither the run time of it nor the end-to-end latency.


Ideally, we wish to leverage the AI accelerator for versatile machine learning tasks (including training) efficiently while ensuring strong security.
Among the above solutions, Graviton is the closest to this goal.
Inspired by Graviton, we will customize AI accelerators in order to extend TEEs to the accelerator for efficient privacy-preserving machine learning.

\section{Methodology}

In this section, we will detail the proposed solution for privacy-preserving machine learning, where there are data owners, model owners and a computing platform.
In this scenario, the data/model owner wishes to use the computing platform (which can be a third party or one of the data owners or the model owners) to accomplish a machine learning task without disclosing their data/model.
To achieve this efficiently, our solution leverages the trusted CPU and the customized AI accelerator inside the computing platform.
After the data owners and the model owners send their data/model into the CPU enclave through encrypted channels, the computation is offloaded to the accelerator securely. 

\begin{figure}[!t]
\centering
\includegraphics[width=\columnwidth]{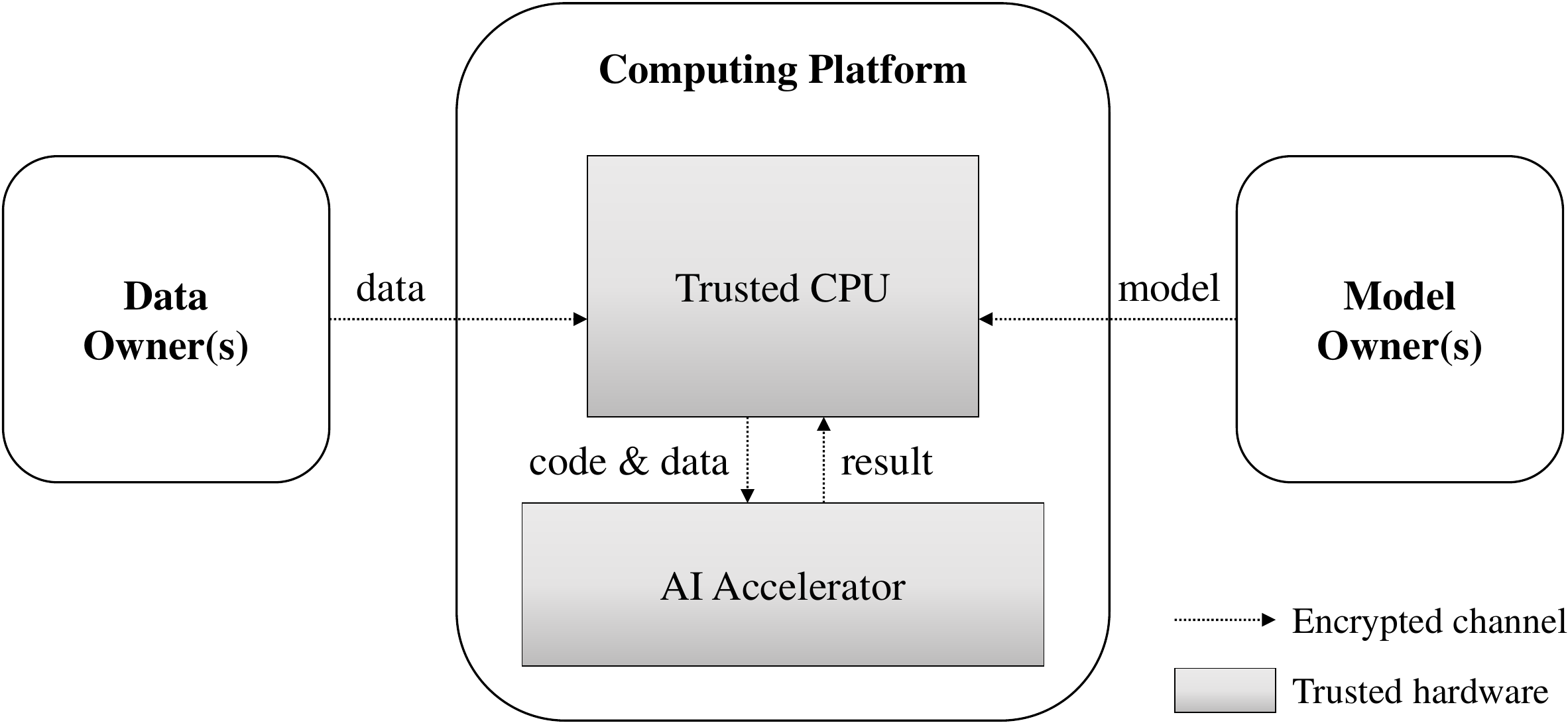}
\caption{High-level overview of our solution for privacy-preserving machine learning.}
\label{fig1}
\end{figure}

The whole procedure is depicted in Figure \ref{fig1}.
In the following part, we will define our threat model first, describe the procedure of trust establishment and our protection method, and then analyze the performance overhead of our solution.

\subsection{Threat Model}
We consider the computing platform can be compromised by an adversary, who can control the entire software/hardware system.

For hardware, only 1) the CPU package and 2) the accelerator package are trusted.
They are considered the only secure regions of the computing platform.
All the other hardware, including memory, storage, peripheral devices, may be compromised.
We consider the adversary can 
\begin{itemize}
\item eavesdrop on the system bus and the PCIe bus
\item directly access the main memory via a malicious device
\item make man-in-the-middle attacks between the CPU and the accelerator
\item eavesdrop on, tamper with or directly access the device memory of the accelerator if the AI accelerator has untrusted device memory (e.g. off-package device memory)
\end{itemize}

For software, everything except the program running in the enclave is untrusted.
Malicious software is able to compromise the operating system or the hypervisor to run at the highest privilege level, so the adversary can
\begin{itemize}
\item directly access any part of the main memory except the PRM (processor's reserved memory) region
\item access and send commands to the accelerator via MMIO
\item compromise the driver and then invoke or tamper with driver APIs if the driver is not running in the enclave\footnote{E.g. Intel SGX does not support kernel-mode enclaves.}
\end{itemize}
 
The adversary wishes to steal the data/model on the computing platform.
The goal of our solution is to prevent the adversary from this and ensure the integrity of code and data.
The size of the data/model, the run time and the memory access pattern are not considered sensitive in this paper.
Since the unmodified trusted CPU is included in our solution, this work will not prevent existing vulnerabilities of the trusted CPU, so side-channel attacks and denial-of-service attacks are out of our scope.

\subsection{Trust}

To establish trust between two parties, there are existing methods in the practice of trusted computing (such as TPM and Intel SGX). 
In our solution, we adopt these methods (generally known as remote attestation \cite{costan16}) to establish trust between the data owner (or the model owner) and the CPU enclave.
As a result, both the data and model owners can trust 1) the enclave is a genuine secure environment provided by the trusted CPU, 2) the attested software is correctly running in this enclave and 3) their data and model have been uploaded to the enclave securely and integrally.

\begin{figure}[t]
\centering
\includegraphics[width=\columnwidth]{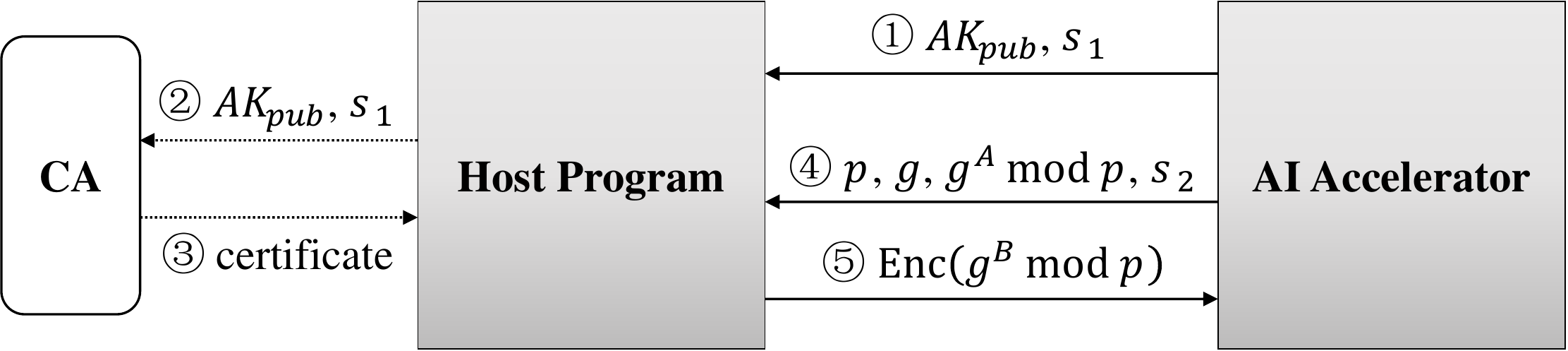}
\caption{Trust establishment between the host program and the AI accelerator.}
\label{fig2}
\end{figure}

Next, to extend the trusted boundary from the CPU enclave to the AI accelerator, we borrow the concept of remote attestation and detail our method depicted in Figure \ref{fig2}.

\subsubsection{Authentication}
The first step is authentication, where the AI accelerator proves itself to the trusted software running in the CPU enclave (denoted by the ``host program'' in the following part) that it is a genuine trusted accelerator. 
The process of authentication is based on public-key cryptography. 
Specifically, each trusted accelerator is assigned with a unique pair of endorsement keys $(EK_{pri}, EK_{pub})$ when it is manufactured.
The private key $EK_{pri}$ is burned to the accelerator hardware and it should be only known by the accelerator.
The public key $EK_{pub}$ is maintained by the manufacturer's certificate authority (CA).
Then, each time the host program wants to use the accelerator, the accelerator generates a different pair of attestation keys $(AK_{pri}, AK_{pub})$, and then sends the public key $AK_{pub}$ and an $EK_{pri}$-signed signature $s_1=\mathrm{Sign}(EK_{pri}, AK_{pub})$ to the host program.

The host program should verify whether $AK_{pub}$ is bound to a genuine trusted accelerator.
For this purpose, it sends the received $AK_{pub}$ and $s_1$ to the manufacturer's CA, and then the CA uses $EK_{pub}$ to verify the signature.
If $s_1$ matches $AK_{pub}$, the CA issues a certificate and returns it to the host program.
Once the host program receives the certificate, it can confirm that $AK_{pub}$ is generated by the trusted accelerator.

\subsubsection{Key exchange}
The second step is key exchange, the purpose of which is to establish a shared symmetric key between the AI accelerator and the host program.
Based on the trust established in the first step, we use AK as the signing key for ephemeral Diffie-Hellman key exchange.
Specifically, the accelerator generates a prime number $p$, a primitive root $g$ and a random number $A$, and then transmits $p$, $g$, $g^A \bmod p$ and an $AK_{pri}$-signed signature $s_2=\mathrm{Sign}(AK_{pri}, p||g||g^A \bmod p)$ to the host program.
After the host program receives them and verifies the signature with $AK_{pub}$, it generates a random number $B$ and then sends $\mathrm{Encrypt}(AK_{pub}, g^B \bmod p)$ to the accelerator.
In the end, both the host program and the accelerator can calculate $g^{AB} \bmod p$ as their shared secret and can derive a symmetric key $K$ from the shared secret by a key derivation function.

\subsection{Protection}
\label{protection}

On the basis of the shared symmetric key, we propose the following data protection methods to prevent the adversary from obtaining the data/model when the host program offloads the computation to the AI accelerator.

\begin{figure*}[h]
\centering
\includegraphics[width=1.75\columnwidth]{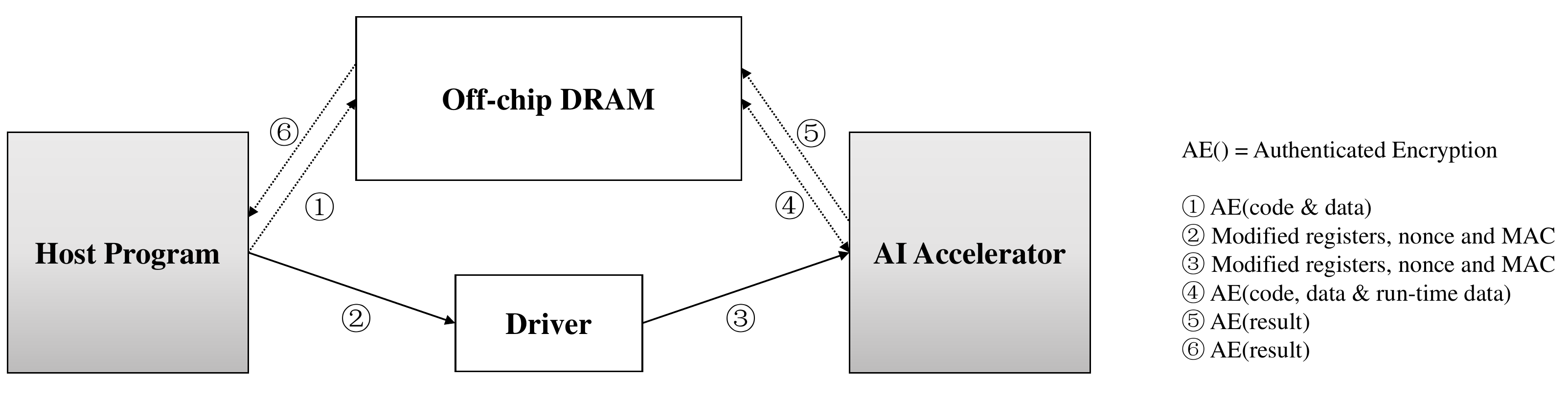}
\caption{Methods for protecting code, data and programmer-visible registers transmitted between the host program and the AI accelerator.}
\label{fig3}
\end{figure*}

\subsubsection{Overview}
First, we protect the code and data transmitted between the host program and the accelerator.
Generally, the code and data are placed from the host program's memory space to the ``off-chip DRAM'', which denotes the memory that is not part of the CPU nor the accelerator package. 
Then, the accelerator fetches the code from the off-chip DRAM and accesses the off-chip DRAM to get/put run-time data.
Considering the adversary can read and write the off-chip DRAM device memory and the buses can be comprised, we should not disclose the code and data in such insecure regions.
Formally, we define:
\begin{principle}
The code and data should be encrypted and integrity-protected during the transmission between the host program and the off-chip DRAM. 
\end{principle}
\begin{principle}
The code and data should keep encrypted and integrity-protected in the off-chip DRAM. 
\end{principle}

Based on these rules, we employ authenticated encryption, which is a combination of encryption and message authentication, to ensure both confidentiality and integrity of the transmitted messages.
In particular, the host program encrypts the code and data in the CPU enclave with the symmetric key $K$, calculates the message authentication code (MAC) of them with $K$ and then writes the ciphertext and the MAC to the off-chip DRAM\footnote{The MAC can be also written to the accelerator's register.}.
To use the code and data, the accelerator decrypts them and verifies their integrity with the MAC.
Symmetrically, after calculating the result, the accelerator encrypts the result and calculates its MAC, and then writes the ciphertext and the MAC to the off-chip DRAM.
The host program can get the result after decryption and verification.

\subsubsection{Detail}
However, to achieve efficiency as well as security, some issues need to be carefully considered:
\begin{itemize}
\item Where is the decrypted code and data placed?
\item How to protect run-time data stored in the off-chip DRAM?
\end{itemize}

If the AI accelerator has its own trusted on-package device memory, the answer is simple: It can just put the decrypted code and data in the trusted device memory and directly store the run-time data without encryption or integrity protection.

If not, the accelerator can never put unencrypted messages in the off-chip DRAM according to Rule 2.
It is also impractical to store them in the accelerator's limited on-chip storage.
Therefore, the accelerator must decrypt the code and data ``on demand''.
In particular, each time the accelerator fetches a piece of code or data from the off-chip DRAM to the accelerator's on-chip storage (e.g. SRAM buffers or registers), it fetches the corresponding piece of ciphertext and decrypts it inside the accelerator.
To achieve this efficiently, we require a counter mode--based authenticated encryption scheme such as AES-GCM and AES-CCM.
Taking AES-GCM for example, after the accelerator fetches the piece of ciphertext, the ciphertext is XORed with $AES(counter)$ inside the accelerator to get the plaintext.
This procedure avoids data dependency and thus it is friendly to hardware implementation.

Integrity is another challenge because the adversary can modify the code and data in the untrusted off-chip DRAM at run time.
Thus, the accelerator needs to verify integrity each time it accesses the off-chip DRAM.
However, it is extremely inefficient to repeatedly verify whether the whole code and data match the MAC.
To tackle this problem, we propose the following scheme: 
\begin{itemize}
\item Before the host program applies authenticated encryption to the code and data, it divides the whole code and data into $m$ pieces, and the size of each piece does not exceed $s$.
\item Then the host program calculates their respective ciphertext and MAC and then stores these $m$ pieces of ciphertext and MAC in the off-chip DRAM.
\item In this case, each time the accelerator accesses the off-chip DRAM, it only fetches corresponding pieces and verifies whether they match their MAC.
\end{itemize}
A smaller $s$ means a finer granularity and causes a lower latency of the cryptographical computation. 
However, it also means a larger $m$ and leads to a larger memory consumption (for the $m$ pieces of MAC) as well as an increase in the accelerator's DRAM access.
As a result, the value of $m$ should be the trade-off between computation and memory access.

The run-time data should also be protected when the AI accelerator has no trusted device memory.
Specifically, according to Rule 2, the run-time data should be encrypted by counter-mode encryption before they are written to the off-chip DRAM, and their integrity should be protected by the method described in the previous paragraph.

\subsubsection{Register state}
So far we have completed the protection of the code and data in the off-chip DRAM to ensure that the adversary cannot learn or tamper with them.
However, the information that is stored in the AI accelerator's programmer-visible registers (e.g. address registers and control registers) has not been protected.
Specifically, if the host program wishes to use the AI accelerator, it invokes the accelerator's driver to write the accelerator's registers through MMIO.
In our threat model, the adversary is able to directly access the registers through MMIO.
Thus, we have the following rule:
\begin{principle}
The integrity and freshness of the programmer-visible registers of the AI accelerator should be guaranteed during the transmission between the host program and the accelerator. 
\end{principle}

Based on this rule, we propose to use message authentication code (MAC) to prevent the adversary from tampering with the registers and use nonces to prevent replay attacks.
Note that the kernel-mode driver may not be running in the CPU enclave, so we let the runtime (also known as the user-mode driver) in the enclave to maintain the register state.
Each time the host program modifies a register, it also calculates the MAC of the whole register state and the nonce with the symmetric key $K$.
After the host program writes the modified registers, it writes the MAC (as well as the nonce) to a specific register of the accelerator.
All the writes are done by invoking the kernel-mode driver.
Since only the host program and the accelerator know $K$, the accelerator can verify the integrity of the register state and the driver cannot counterfeit the MAC.
Therefore, this problem is solved even if the driver is not within the trust boundary (i.e. the driver is untrusted).
To sum up, Figure \ref{fig3} depicts our protection methods.

\subsection{Overhead}
\label{overhead}
Analytically, the total run time of a machine learning task is composed of computation and memory access.
Compared with conventional AI accelerators, our methods do not affect the computation (e.g. matrix multiplications).
In contrast, our protection methods additionally introduce encryption, decryption, and message authentication, all of which are bound to memory access.
Each time the trusted accelerator fetches a piece of code/data from the off-chip DRAM, the accelerator has to decrypt it and verify its MAC before using it.
Each time the trusted accelerator aims to write a piece of data to the off-chip DRAM, the accelerator has to encrypt it and calculate its MAC.
As a result, the slowdown will be significant if the workload involves a large amount of memory access.
On the contrary, a small amount of memory access will lead to slight slowdowns.
Therefore, the extent of the performance overhead depends on the memory access intensity of the workload.


\section{Case Study: VTA}
\label{casestudy}
In this section, we demonstrate the process of customizing a trusted AI accelerator using the proposed methods. 
In particular, we implement the customization on the open-source versatile tensor accelerator (VTA) \cite{moreau19} and evaluate its impact on the performance of the VTA.

\subsection{Architecture}
Referring to Figure \ref{fig1}, we deploy a trusted CPU and a customized VTA in the computing platform, and we have the model/data owners upload their private model and data to a CPU enclave securely.
Then, the host program (a trusted and attested piece of software running in the enclave) compiles the model into VTA code just in time with the TVM deep learning compilation stack \cite{chen18}, and then offloads the code and data to the VTA and waits until the result is calculated following the proposed protection methods.
To implement support for the protection methods as well as the trust establishment methods on the VTA, both hardware and software of the VTA are modified.

\begin{figure}[!t]
\centering
\includegraphics[width=\columnwidth]{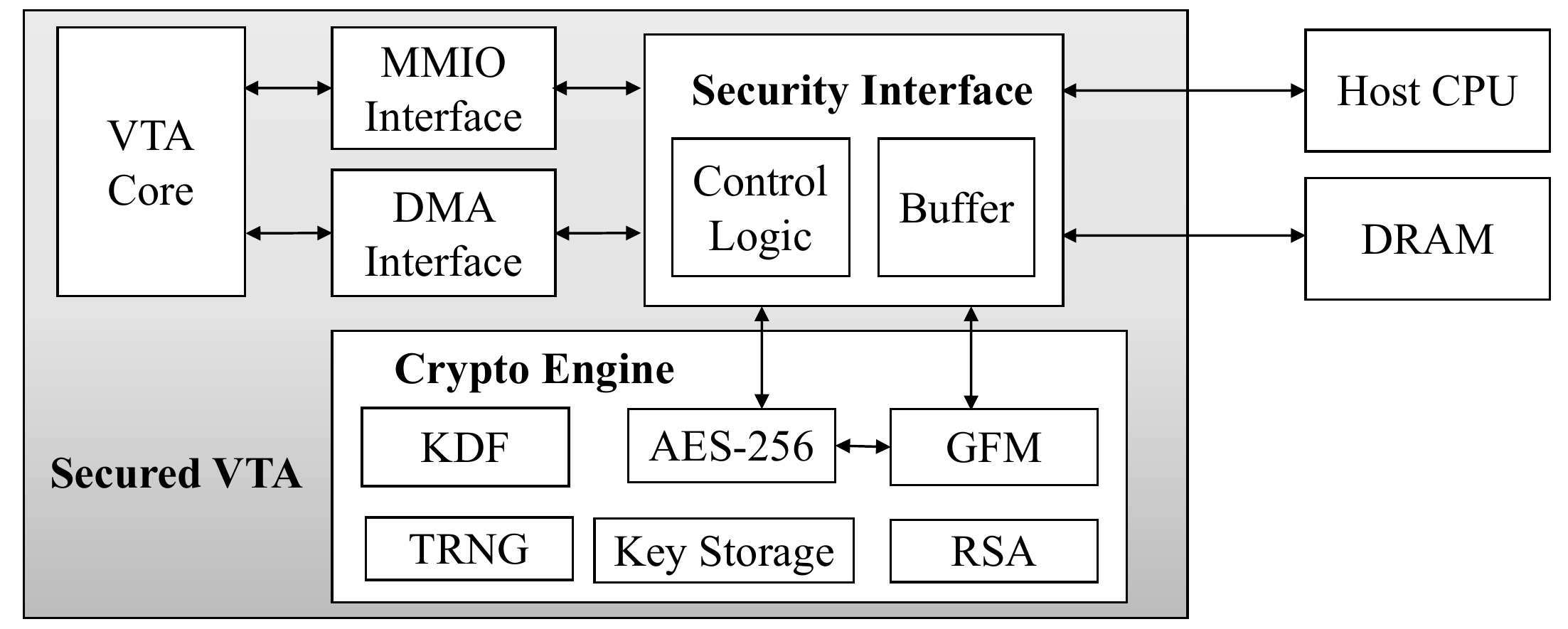}
\caption{We customize the VTA by adding a security interface and a crypto engine.}

\label{fig4}
\end{figure}

In terms of hardware, we do not change the design of the VTA core\footnote{https://github.com/apache/incubator-tvm/tree/v0.6/vta}.
Instead, a security layer, including a security interface and a crypto engine, is added between the VTA core and the host CPU/DRAM (Figure \ref{fig4}) to handle the cryptographic functionalities.
The security interface buffers the received data in an on-chip buffer (2~KB), communicates the data to the crypto engine, and then sends the encrypted/decrypted data to the DRAM/VTA.
The crypto engine contains components (such as AES, RSA, and TRNG) that are used for authenticated encryption and trust establishment.
In this experiment, we implement the AES-256 module that employs a pipelined structure in which encryption/decryption of a 128-bit plaintext/ciphertext takes 29 clock cycles\footnote{https://opencores.org/projects/tiny\_aes}, and the GFM (Galois-Field Multiplication) module that incurs 8 clock cycles for the authentication of each 128-bit text\footnote{https://opencores.org/projects/gcm-aes}.

The modification of software only involves the VTA runtime, while the instruction set architecture of VTA or the just-in-time compiler does not need to change in our solution. 
The VTA runtime (\texttt{vta/src/runtime.cc}) is part of the host program which acts as an interface to the untrusted off-chip DRAM and the kernel-mode driver.
Specifically, the original VTA runtime is responsible for allocating a few buffers in the off-chip DRAM, putting the code (namely ``kernel'' in VTA) and data into corresponding buffers, and invoking the driver to launch the kernel.
In our solution, the runtime is modified such that it can 1) apply authenticated encryption to the code and data, and 2) maintain the register state to calculate MAC of the registers and nonces.
The memory layout remains the same because counter-mode encryption does not change the size of messages, but authenticated encryption results in extra metadata (i.e., nonces and GMAC) in addition to the original message.
The metadata is stored in a newly-allocated buffer in the off-chip DRAM.


\subsection{Evaluation}
We build a cycle-accurate simulation environment based on VTA's TSIM\footnote{https://github.com/apache/incubator-tvm/tree/v0.6/vta/apps/tsim\_example}, which can compile the register-transfer-level design into \texttt{libvta\_hw.so} and integrate this simulation library into the TVM stack to evaluate run time in clock cycles.

Since the VTA is mainly designed for computer-vision tasks \cite{moreau19}, we first test the computation of 2D convolution layers and fully connected layers, which represent the dominant operators in most computer vision deep neural networks. 
Specifically, we use two convolutional layers and two fully connected layers of AlexNet \cite{krizhevsky14}, which have similar amount of computation, as our benchmarks: Conv4 (384 input channels, 256 output channels), Conv5 (256 input channels, 256 output channels), FC1 (9216 inputs, 4096 outputs) and FC2 (4096 inputs, 4096 outputs).
For each benchmark, the run time of three configurations are evaluated: 1) the original VTA without any protection (denoted by VTA), 2) the VTA with full protection (including confidentiality, integrity and freshness, denoted by VTA-trusted), and 3) the VTA with the protection of only confidentiality (implemented by AES-CTR, denoted by VTA-ctr), which is used for profiling.

\begin{table}[!t]
\caption{Latency in clock cycles (and slowdowns compared with the original VTA).}
\label{tab1}
\resizebox{\columnwidth}{!}{
\centering
\begin{tabular}{cccc}
\hline
\textbf{} & VTA & VTA-trusted & VTA-ctr \\ \hline
Conv4 & 2~782~962 & 2~988~247 (1.074$\times$) & 2~872~727 (1.032$\times$) \\
Conv5 & 1~879~117& 2~083~659 (1.109$\times$) & 1~969~399 (1.048$\times$)  \\
FC1 & 5~418~983 & 29~300~635 (5.407$\times$) & 6~016~817 (1.110$\times$) \\
FC2 & 2~412~609 & 13~034~043 (5.402$\times$) & 2~682~866 (1.112$\times$) \\ \hline
ResNet-18 & 29~964~469 & 32~338~145 (1.079$\times$) & 30~238~890 (1.009$\times$) \\ \hline
\end{tabular}
}
\end{table}

The experimental result is shown in Table \ref{tab1}.
Compared with the original VTA, VTA-trusted results in 1.074$\times$--1.109$\times$ slowdowns for convolutional layers but results in 5.402$\times$--5.407$\times$ slowdowns with respect to fully connected layers.
It is noticed that the slowdowns with respect to different types of layers vary significantly. 
We will explain this as follows.

There are two main factors that lead to the significant overhead of computing fully connected layers.
The first one is memory access intensity.
As we discussed in Section \ref{overhead}, the extent of the performance overhead depends on the memory access intensity of the workload.
For a fully connected layer (with only 1 batch), the dominant operator is vector-matrix multiplication.
However, the memory access intensity of the vector-matrix multiplication is about 1 word/FLOP, which is notably higher than 2D convolution (approximately $\frac{1}{H_o\times W_o}$ word/FLOP).

The second factor stems from the implementation of the GFM module which is used for calculating GMAC.
Specifically, we tested a configuration without the GFM module (i.e. VTA-ctr) for profiling.
As the result shows, removing the GFM module results in a much better performance for fully connected layers compared with VTA-trusted.
The slowdowns against the original VTA are reduced from 5.402$\times$--5.407$\times$ to 1.110$\times$--1.112$\times$.
We notice that the implementation of the AES module is pipelined and counter-mode encryption avoids data dependency, so it is feasible to achieve high throughput.
As a result, VTA-ctr only incurs 1.032$\times$--1.112$\times$ slowdowns.
However, our implementation of the GFM module is not pipelined and the calculation of GMAC has strong data dependency.
Thus, authenticating a piece of data of $s$ bits needs $\lceil\frac{s}{128}\rceil\times 8$ clock cycles in our implementation.

\subsection{Potential Improvement}
Based on the observations above, we propose several potential ways to improve the trusted VTA in order to reduce slowdowns.
\begin{itemize}
    \item To reduce the memory access intensity, we can optimize the machine learning model at algorithm level, such as reducing/avoiding fully connected layers, increasing the batch size, using sparse neural networks, etc.
    \item To reduce the memory access intensity, we may also optimize the computation at the compiler level. For example, recent deep learning compilers are exploiting dedicated optimization to achieve efficient model-to-hardware mapping (e.g. TVM). 
    \item To reduce the overhead of message authentication, optimizing hardware implementation of the authentication module (e.g. designing a low-latency GFM module \cite{lu_2009}) is a straightforward way. 
    \item To reduce the overhead of message authentication, using other authentication schemes may be a potential direction. For example, the Merkle tree can reduce the computation complexity of authenticating $s$-bit data  from $O(s)$ to $O(\log s)$ because of its potential for parallelism. Theoretically, the best effect (i.e. upperbound) that this method can achieve is the result of VTA-ctr, which has been shown in Table \ref{tab1}. 
\end{itemize}

Fortunately, the current version of TVM (v0.6) has supported end-to-end compilation for ResNet-18 \cite{he16} on VTA with compilation optimization (support for other models is still under development).
Thus, we take ResNet-18 as an additional benchmark, whose result is shown in Table \ref{tab1}.
As we can see, the slowdown of VTA-trusted is satisfactorily small (1.079$\times$), which is attributed to the low memory access intensity after compilation optimization.
In addition, the slowdown of VTA-ctr (1.009$\times$) indicates a large space for accelerating message authentication.

\section{Conclusion}
This paper proposes an efficient solution for privacy-preserving machine learning with AI accelerators.
The key innovation lies in incorporating a customized AI accelerator into the computing platform and ensuring its security.
With the help of the proposed methods, the users can upload their private model and data to the computing platform securely and calculates the result efficiently within the AI accelerator.
The case study shows this solution is effective on the versatile tensor accelerator at a small design cost and  moderate performance overhead, and it is promising to use this solution in the industry with lower overhead.
\bibliographystyle{ACM-Reference-Format}
\bibliography{references}


\begin{thebibliography}{22}


\ifx \showCODEN    \undefined \def \showCODEN     #1{\unskip}     \fi
\ifx \showDOI      \undefined \def \showDOI       #1{#1}\fi
\ifx \showISBNx    \undefined \def \showISBNx     #1{\unskip}     \fi
\ifx \showISBNxiii \undefined \def \showISBNxiii  #1{\unskip}     \fi
\ifx \showISSN     \undefined \def \showISSN      #1{\unskip}     \fi
\ifx \showLCCN     \undefined \def \showLCCN      #1{\unskip}     \fi
\ifx \shownote     \undefined \def \shownote      #1{#1}          \fi
\ifx \showarticletitle \undefined \def \showarticletitle #1{#1}   \fi
\ifx \showURL      \undefined \def \showURL       {\relax}        \fi
\providecommand\bibfield[2]{#2}
\providecommand\bibinfo[2]{#2}
\providecommand\natexlab[1]{#1}
\providecommand\showeprint[2][]{arXiv:#2}

\bibitem[\protect\citeauthoryear{Chen, Moreau, Jiang, Zheng, Yan, Shen, Cowan,
  Wang, Hu, Ceze, Guestrin, and Krishnamurthy}{Chen et~al\mbox{.}}{2018}]%
        {chen18}
\bibfield{author}{\bibinfo{person}{Tianqi Chen}, \bibinfo{person}{Thierry
  Moreau}, \bibinfo{person}{Ziheng Jiang}, \bibinfo{person}{Lianmin Zheng},
  \bibinfo{person}{Eddie~Q. Yan}, \bibinfo{person}{Haichen Shen},
  \bibinfo{person}{Meghan Cowan}, \bibinfo{person}{Leyuan Wang},
  \bibinfo{person}{Yuwei Hu}, \bibinfo{person}{Luis Ceze},
  \bibinfo{person}{Carlos Guestrin}, {and} \bibinfo{person}{Arvind
  Krishnamurthy}.} \bibinfo{year}{2018}\natexlab{}.
\newblock \showarticletitle{{TVM}: {An} {Automated} {End}-to-{End} {Optimizing}
  {Compiler} for {Deep} {Learning}}. In \bibinfo{booktitle}{\emph{{USENIX}
  {Symposium} on {Operating} {Systems} {Design} and {Implementation}
  ({OSDI})}}. \bibinfo{pages}{578--594}.
\newblock


\bibitem[\protect\citeauthoryear{Chen, Emer, and Sze}{Chen
  et~al\mbox{.}}{2016}]%
        {chen16}
\bibfield{author}{\bibinfo{person}{Yu-Hsin Chen}, \bibinfo{person}{Joel~S.
  Emer}, {and} \bibinfo{person}{Vivienne Sze}.}
  \bibinfo{year}{2016}\natexlab{}.
\newblock \showarticletitle{Eyeriss: {A} {Spatial} {Architecture} for
  {Energy}-{Efficient} {Dataflow} for {Convolutional} {Neural} {Networks}}. In
  \bibinfo{booktitle}{\emph{International {Symposium} on {Computer}
  {Architecture} ({ISCA})}}. \bibinfo{pages}{367--379}.
\newblock


\bibitem[\protect\citeauthoryear{Corporation}{Corporation}{2016}]%
        {intel_2016}
\bibfield{author}{\bibinfo{person}{Intel Corporation}.}
  \bibinfo{year}{2016}\natexlab{}.
\newblock \bibinfo{booktitle}{\emph{Intel® 64 and {IA}-32 {Architectures}
  {Software} {Developer} {Manuals}}}.
\newblock


\bibitem[\protect\citeauthoryear{Costan and Devadas}{Costan and
  Devadas}{2016}]%
        {costan16}
\bibfield{author}{\bibinfo{person}{Victor Costan} {and}
  \bibinfo{person}{Srinivas Devadas}.} \bibinfo{year}{2016}\natexlab{}.
\newblock \showarticletitle{Intel {SGX} {Explained}}.
\newblock \bibinfo{journal}{\emph{IACR Cryptology ePrint Archive}}
  \bibinfo{volume}{2016} (\bibinfo{year}{2016}), \bibinfo{pages}{86}.
\newblock


\bibitem[\protect\citeauthoryear{Gilad-Bachrach, Dowlin, Laine, Lauter,
  Naehrig, and Wernsing}{Gilad-Bachrach et~al\mbox{.}}{2016}]%
        {gilad-bachrach16}
\bibfield{author}{\bibinfo{person}{Ran Gilad-Bachrach}, \bibinfo{person}{Nathan
  Dowlin}, \bibinfo{person}{Kim Laine}, \bibinfo{person}{Kristin~E. Lauter},
  \bibinfo{person}{Michael Naehrig}, {and} \bibinfo{person}{John Wernsing}.}
  \bibinfo{year}{2016}\natexlab{}.
\newblock \showarticletitle{{CryptoNets}: {Applying} {Neural} {Networks} to
  {Encrypted} {Data} with {High} {Throughput} and {Accuracy}}. In
  \bibinfo{booktitle}{\emph{International {Conference} on {Machine} {Learning}
  ({ICML})}}, Vol.~\bibinfo{volume}{48}. \bibinfo{pages}{201--210}.
\newblock


\bibitem[\protect\citeauthoryear{He, Zhang, Ren, and Sun}{He
  et~al\mbox{.}}{2016}]%
        {he16}
\bibfield{author}{\bibinfo{person}{Kaiming He}, \bibinfo{person}{Xiangyu
  Zhang}, \bibinfo{person}{Shaoqing Ren}, {and} \bibinfo{person}{Jian Sun}.}
  \bibinfo{year}{2016}\natexlab{}.
\newblock \showarticletitle{Deep {Residual} {Learning} for {Image}
  {Recognition}}. In \bibinfo{booktitle}{\emph{Conference on {Computer}
  {Vision} and {Pattern} {Recognition} ({CVPR})}}. \bibinfo{pages}{770--778}.
\newblock


\bibitem[\protect\citeauthoryear{Hunt, Song, Shokri, Shmatikov, and
  Witchel}{Hunt et~al\mbox{.}}{2018}]%
        {hunt18}
\bibfield{author}{\bibinfo{person}{Tyler Hunt}, \bibinfo{person}{Congzheng
  Song}, \bibinfo{person}{Reza Shokri}, \bibinfo{person}{Vitaly Shmatikov},
  {and} \bibinfo{person}{Emmett Witchel}.} \bibinfo{year}{2018}\natexlab{}.
\newblock \showarticletitle{Chiron: {Privacy}-{Preserving} {Machine} {Learning}
  as a {Service}}.
\newblock \bibinfo{journal}{\emph{CoRR}}  \bibinfo{volume}{abs/1803.05961}
  (\bibinfo{year}{2018}).
\newblock


\bibitem[\protect\citeauthoryear{Jang, Tang, Kim, Sethumadhavan, and Huh}{Jang
  et~al\mbox{.}}{2019}]%
        {jang19}
\bibfield{author}{\bibinfo{person}{Insu Jang}, \bibinfo{person}{Adrian Tang},
  \bibinfo{person}{Taehoon Kim}, \bibinfo{person}{Simha Sethumadhavan}, {and}
  \bibinfo{person}{Jaehyuk Huh}.} \bibinfo{year}{2019}\natexlab{}.
\newblock \showarticletitle{Heterogeneous {Isolated} {Execution} for
  {Commodity} {GPUs}}. In \bibinfo{booktitle}{\emph{International {Conference}
  on {Architectural} {Support} for {Programming} {Languages} and {Operating}
  {Systems} ({ASPLOS})}}. \bibinfo{pages}{455--468}.
\newblock


\bibitem[\protect\citeauthoryear{Jouppi, Young, Patil, Patterson, Agrawal,
  Bajwa, Bates, Bhatia, Boden, Borchers, Boyle, Cantin, Chao, Clark, Coriell,
  Daley, Dau, Dean, Gelb, Ghaemmaghami, Gottipati, Gulland, Hagmann, Ho,
  Hogberg, Hu, Hundt, Hurt, Ibarz, Jaffey, Jaworski, Kaplan, Khaitan,
  Killebrew, Koch, Kumar, Lacy, Laudon, Law, Le, Leary, Liu, Lucke, Lundin,
  MacKean, Maggiore, Mahony, Miller, Nagarajan, Narayanaswami, Ni, Nix, Norrie,
  Omernick, Penukonda, Phelps, Ross, Ross, Salek, Samadiani, Severn, Sizikov,
  Snelham, Souter, Steinberg, Swing, Tan, Thorson, Tian, Toma, Tuttle,
  Vasudevan, Walter, Wang, Wilcox, and Yoon}{Jouppi et~al\mbox{.}}{2017}]%
        {jouppi17}
\bibfield{author}{\bibinfo{person}{Norman~P. Jouppi}, \bibinfo{person}{Cliff
  Young}, \bibinfo{person}{Nishant Patil}, \bibinfo{person}{David~A.
  Patterson}, \bibinfo{person}{Gaurav Agrawal}, \bibinfo{person}{Raminder
  Bajwa}, \bibinfo{person}{Sarah Bates}, \bibinfo{person}{Suresh Bhatia},
  \bibinfo{person}{Nan Boden}, \bibinfo{person}{Al Borchers},
  \bibinfo{person}{Rick Boyle}, \bibinfo{person}{Pierre-luc Cantin},
  \bibinfo{person}{Clifford Chao}, \bibinfo{person}{Chris Clark},
  \bibinfo{person}{Jeremy Coriell}, \bibinfo{person}{Mike Daley},
  \bibinfo{person}{Matt Dau}, \bibinfo{person}{Jeffrey Dean},
  \bibinfo{person}{Ben Gelb}, \bibinfo{person}{Tara~Vazir Ghaemmaghami},
  \bibinfo{person}{Rajendra Gottipati}, \bibinfo{person}{William Gulland},
  \bibinfo{person}{Robert Hagmann}, \bibinfo{person}{C.~Richard Ho},
  \bibinfo{person}{Doug Hogberg}, \bibinfo{person}{John Hu},
  \bibinfo{person}{Robert Hundt}, \bibinfo{person}{Dan Hurt},
  \bibinfo{person}{Julian Ibarz}, \bibinfo{person}{Aaron Jaffey},
  \bibinfo{person}{Alek Jaworski}, \bibinfo{person}{Alexander Kaplan},
  \bibinfo{person}{Harshit Khaitan}, \bibinfo{person}{Daniel Killebrew},
  \bibinfo{person}{Andy Koch}, \bibinfo{person}{Naveen Kumar},
  \bibinfo{person}{Steve Lacy}, \bibinfo{person}{James Laudon},
  \bibinfo{person}{James Law}, \bibinfo{person}{Diemthu Le},
  \bibinfo{person}{Chris Leary}, \bibinfo{person}{Zhuyuan Liu},
  \bibinfo{person}{Kyle Lucke}, \bibinfo{person}{Alan Lundin},
  \bibinfo{person}{Gordon MacKean}, \bibinfo{person}{Adriana Maggiore},
  \bibinfo{person}{Maire Mahony}, \bibinfo{person}{Kieran Miller},
  \bibinfo{person}{Rahul Nagarajan}, \bibinfo{person}{Ravi Narayanaswami},
  \bibinfo{person}{Ray Ni}, \bibinfo{person}{Kathy Nix},
  \bibinfo{person}{Thomas Norrie}, \bibinfo{person}{Mark Omernick},
  \bibinfo{person}{Narayana Penukonda}, \bibinfo{person}{Andy Phelps},
  \bibinfo{person}{Jonathan Ross}, \bibinfo{person}{Matt Ross},
  \bibinfo{person}{Amir Salek}, \bibinfo{person}{Emad Samadiani},
  \bibinfo{person}{Chris Severn}, \bibinfo{person}{Gregory Sizikov},
  \bibinfo{person}{Matthew Snelham}, \bibinfo{person}{Jed Souter},
  \bibinfo{person}{Dan Steinberg}, \bibinfo{person}{Andy Swing},
  \bibinfo{person}{Mercedes Tan}, \bibinfo{person}{Gregory Thorson},
  \bibinfo{person}{Bo Tian}, \bibinfo{person}{Horia Toma},
  \bibinfo{person}{Erick Tuttle}, \bibinfo{person}{Vijay Vasudevan},
  \bibinfo{person}{Richard Walter}, \bibinfo{person}{Walter Wang},
  \bibinfo{person}{Eric Wilcox}, {and} \bibinfo{person}{Doe~Hyun Yoon}.}
  \bibinfo{year}{2017}\natexlab{}.
\newblock \showarticletitle{In-{Datacenter} {Performance} {Analysis} of a
  {Tensor} {Processing} {Unit}}. In \bibinfo{booktitle}{\emph{International
  {Symposium} on {Computer} {Architecture} ({ISCA})}}. \bibinfo{pages}{1--12}.
\newblock


\bibitem[\protect\citeauthoryear{Juvekar, Vaikuntanathan, and
  Chandrakasan}{Juvekar et~al\mbox{.}}{2018}]%
        {juvekar18}
\bibfield{author}{\bibinfo{person}{Chiraag Juvekar}, \bibinfo{person}{Vinod
  Vaikuntanathan}, {and} \bibinfo{person}{Anantha Chandrakasan}.}
  \bibinfo{year}{2018}\natexlab{}.
\newblock \showarticletitle{{GAZELLE}: {A} {Low} {Latency} {Framework} for
  {Secure} {Neural} {Network} {Inference}}. In
  \bibinfo{booktitle}{\emph{{USENIX} {Security} {Symposium}}}.
  \bibinfo{pages}{1651--1669}.
\newblock


\bibitem[\protect\citeauthoryear{Krizhevsky}{Krizhevsky}{2014}]%
        {krizhevsky14}
\bibfield{author}{\bibinfo{person}{Alex Krizhevsky}.}
  \bibinfo{year}{2014}\natexlab{}.
\newblock \showarticletitle{One {Weird} {Trick} for {Parallelizing}
  {Convolutional} {Neural} {Networks}}.
\newblock \bibinfo{journal}{\emph{CoRR}}  \bibinfo{volume}{abs/1404.5997}
  (\bibinfo{year}{2014}).
\newblock


\bibitem[\protect\citeauthoryear{Lee, Kohlbrenner, Shinde, Asanovic, and
  Song}{Lee et~al\mbox{.}}{2020}]%
        {lee_2020}
\bibfield{author}{\bibinfo{person}{Dayeol Lee}, \bibinfo{person}{David
  Kohlbrenner}, \bibinfo{person}{Shweta Shinde}, \bibinfo{person}{Krste
  Asanovic}, {and} \bibinfo{person}{Dawn Song}.}
  \bibinfo{year}{2020}\natexlab{}.
\newblock \showarticletitle{Keystone: {An} {Open} {Framework} for
  {Architecting} {Trusted} {Execution} {Environments}}. In
  \bibinfo{booktitle}{\emph{European {Conference} on {Computer} {Systems}
  ({EuroSys})}}.
\newblock


\bibitem[\protect\citeauthoryear{Liang, Li, Chen, Jiang, Xie, and Yang}{Liang
  et~al\mbox{.}}{2020}]%
        {liang_2020}
\bibfield{author}{\bibinfo{person}{Hongliang Liang}, \bibinfo{person}{Mingyu
  Li}, \bibinfo{person}{Yixiu Chen}, \bibinfo{person}{Lin Jiang},
  \bibinfo{person}{Zhuosi Xie}, {and} \bibinfo{person}{Tianqi Yang}.}
  \bibinfo{year}{2020}\natexlab{}.
\newblock \showarticletitle{Establishing {Trusted} {I}/{O} {Paths} for {SGX}
  {Client} {Systems} {With} {Aurora}}.
\newblock \bibinfo{journal}{\emph{IEEE Transactions on Information Forensics
  and Security}}  \bibinfo{volume}{15} (\bibinfo{year}{2020}),
  \bibinfo{pages}{1589--1600}.
\newblock


\bibitem[\protect\citeauthoryear{Limited}{Limited}{2009}]%
        {arm_2009}
\bibfield{author}{\bibinfo{person}{Arm Limited}.}
  \bibinfo{year}{2009}\natexlab{}.
\newblock \bibinfo{booktitle}{\emph{{ARM} {Security} {Technology} {Building} a
  {Secure} {System} using {TrustZone} {Technology}}}.
\newblock


\bibitem[\protect\citeauthoryear{Lu, Shou, Hu, and Guo}{Lu
  et~al\mbox{.}}{2009}]%
        {lu_2009}
\bibfield{author}{\bibinfo{person}{Yang Lu}, \bibinfo{person}{Guochu Shou},
  \bibinfo{person}{Yihong Hu}, {and} \bibinfo{person}{Zhigang Guo}.}
  \bibinfo{year}{2009}\natexlab{}.
\newblock \showarticletitle{The {Research} and {Efficient} {FPGA}
  {Implementation} of {Ghash} {Core} for {GMAC}}. In
  \bibinfo{booktitle}{\emph{International {Conference} on {E}-{Business} and
  {Information} {System} {Security} ({EBISS})}}. \bibinfo{pages}{1--5}.
\newblock


\bibitem[\protect\citeauthoryear{Mohassel and Zhang}{Mohassel and
  Zhang}{2017}]%
        {mohassel17}
\bibfield{author}{\bibinfo{person}{Payman Mohassel} {and}
  \bibinfo{person}{Yupeng Zhang}.} \bibinfo{year}{2017}\natexlab{}.
\newblock \showarticletitle{{SecureML}: {A} {System} for {Scalable}
  {Privacy}-{Preserving} {Machine} {Learning}}. In
  \bibinfo{booktitle}{\emph{{IEEE} {Symposium} on {Security} and {Privacy}
  ({S}\&{P})}}. \bibinfo{pages}{19--38}.
\newblock


\bibitem[\protect\citeauthoryear{Moreau, Chen, Vega, Roesch, Yan, Zheng, Fromm,
  Jiang, Ceze, Guestrin, and Krishnamurthy}{Moreau et~al\mbox{.}}{2019}]%
        {moreau19}
\bibfield{author}{\bibinfo{person}{Thierry Moreau}, \bibinfo{person}{Tianqi
  Chen}, \bibinfo{person}{Luis Vega}, \bibinfo{person}{Jared Roesch},
  \bibinfo{person}{Eddie~Q. Yan}, \bibinfo{person}{Lianmin Zheng},
  \bibinfo{person}{Josh Fromm}, \bibinfo{person}{Ziheng Jiang},
  \bibinfo{person}{Luis Ceze}, \bibinfo{person}{Carlos Guestrin}, {and}
  \bibinfo{person}{Arvind Krishnamurthy}.} \bibinfo{year}{2019}\natexlab{}.
\newblock \showarticletitle{A {Hardware}-{Software} {Blueprint} for {Flexible}
  {Deep} {Learning} {Specialization}}.
\newblock \bibinfo{journal}{\emph{IEEE Micro}} \bibinfo{volume}{39},
  \bibinfo{number}{5} (\bibinfo{year}{2019}), \bibinfo{pages}{8--16}.
\newblock


\bibitem[\protect\citeauthoryear{Ohrimenko, Schuster, Fournet, Mehta, Nowozin,
  Vaswani, and Costa}{Ohrimenko et~al\mbox{.}}{2016}]%
        {ohrimenko16}
\bibfield{author}{\bibinfo{person}{Olga Ohrimenko}, \bibinfo{person}{Felix
  Schuster}, \bibinfo{person}{Cédric Fournet}, \bibinfo{person}{Aastha Mehta},
  \bibinfo{person}{Sebastian Nowozin}, \bibinfo{person}{Kapil Vaswani}, {and}
  \bibinfo{person}{Manuel Costa}.} \bibinfo{year}{2016}\natexlab{}.
\newblock \showarticletitle{Oblivious {Multi}-{Party} {Machine} {Learning} on
  {Trusted} {Processors}}. In \bibinfo{booktitle}{\emph{{USENIX} {Security}
  {Symposium}}}. \bibinfo{pages}{619--636}.
\newblock


\bibitem[\protect\citeauthoryear{Rouhani, Riazi, and Koushanfar}{Rouhani
  et~al\mbox{.}}{2018}]%
        {rouhani18}
\bibfield{author}{\bibinfo{person}{Bita~Darvish Rouhani},
  \bibinfo{person}{M.~Sadegh Riazi}, {and} \bibinfo{person}{Farinaz
  Koushanfar}.} \bibinfo{year}{2018}\natexlab{}.
\newblock \showarticletitle{Deepsecure: {Scalable} {Provably}-{Secure} {Deep}
  {Learning}}. In \bibinfo{booktitle}{\emph{Design {Automation} {Conference}
  ({DAC})}}. \bibinfo{pages}{2:1--2:6}.
\newblock


\bibitem[\protect\citeauthoryear{Tramèr and Boneh}{Tramèr and Boneh}{2019}]%
        {tramer19}
\bibfield{author}{\bibinfo{person}{Florian Tramèr} {and} \bibinfo{person}{Dan
  Boneh}.} \bibinfo{year}{2019}\natexlab{}.
\newblock \showarticletitle{Slalom: {Fast}, {Verifiable} and {Private}
  {Execution} of {Neural} {Networks} in {Trusted} {Hardware}}. In
  \bibinfo{booktitle}{\emph{International {Conference} on {Learning}
  {Representations} ({ICLR})}}.
\newblock


\bibitem[\protect\citeauthoryear{Volos, Vaswani, and Bruno}{Volos
  et~al\mbox{.}}{2018}]%
        {volos18}
\bibfield{author}{\bibinfo{person}{Stavros Volos}, \bibinfo{person}{Kapil
  Vaswani}, {and} \bibinfo{person}{Rodrigo Bruno}.}
  \bibinfo{year}{2018}\natexlab{}.
\newblock \showarticletitle{Graviton: {Trusted} {Execution} {Environments} on
  {GPUs}}. In \bibinfo{booktitle}{\emph{{USENIX} {Symposium} on {Operating}
  {Systems} {Design} and {Implementation} ({OSDI})}}.
  \bibinfo{pages}{681--696}.
\newblock


\bibitem[\protect\citeauthoryear{Weiser and Werner}{Weiser and Werner}{2017}]%
        {weiser_2017}
\bibfield{author}{\bibinfo{person}{Samuel Weiser} {and} \bibinfo{person}{Mario
  Werner}.} \bibinfo{year}{2017}\natexlab{}.
\newblock \showarticletitle{{SGXIO}: {Generic} {Trusted} {I}/{O} {Path} for
  {Intel} {SGX}}. In \bibinfo{booktitle}{\emph{Conference on {Data} and
  {Application} {Security} and {Privacy} ({CODASPY})}}.
  \bibinfo{pages}{261--268}.
\newblock


\end{thebibliography}

\end{document}